**Self-Assembly of Crowded Semiflexible Polymers under Dynamic and Deformable Confinement**

Nasir Amiri[1], Jonathan P. Singer[2], Xin Yong[1]*

[1]Department of Mechanical and Aerospace Engineering, University at Buffalo, Buffalo 14260, USA

[2]Department of Mechanical and Aerospace Engineering, Rutgers University, Piscataway, NJ 08854, USA

*Email: xinyong@buffalo.edu

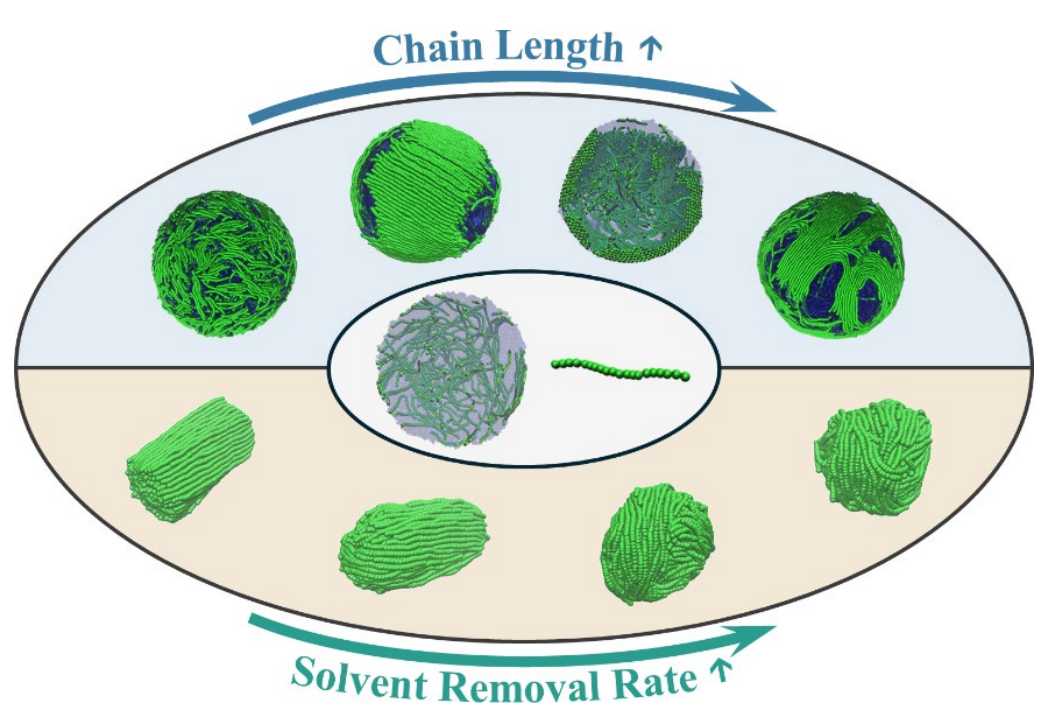

For Table of Contents use only

## Abstract

Semiflexible polymers are ubiquitous in natural and artificial systems, where their intermediate rigidity gives rise to rich structural and dynamical behavior. Confinement plays a central role in these behaviors, as spatial restrictions can promote chain alignment, induce structural rearrangements, and enable complex self-assembly. While the organization of semiflexible polymers under rigid confinement has been extensively investigated, their behavior within deformable and dynamically evolving microenvironments, such as drying droplets or intracellular compartments, remains poorly understood. In this study, we use dissipative particle dynamics simulations to investigate the self-assembly of crowded semiflexible polymers confined within a deformable droplet, whose size may also change over time. By systematically varying polymer contour length, concentration, and degree of confinement, we identify distinct assembly regimes. Increasing polymer concentration promotes the formation of ordered fibrillar domains, with orientational alignment strongest near the droplet interface. Chain length critically dictates the morphology of assembled structures: short chains remain largely disordered, chains with intermediate lengths form linear fibrillar structures with maximal nematic order, and long chains assemble into circular bundles. Dynamic confinement further modulates the assembly through the competition between the rate of confinement change and polymer mobility. Slow increase in the degree of confinement allows polymers to reorganize into highly ordered structures, while rapid crowding kinetically traps the system in disordered states. Our findings elucidate how polymer mechanics and time-dependent confinement jointly govern the organization of semiflexible polymers in deformable, dynamic, and crowded environments.

## Introduction

Polymers are vital materials that constitute numerous biological systems and underpin broad technological applications.[1-5] They possess a wide spectrum of flexibility, which significantly governs their structural and dynamic properties.[6, 7] The theoretical framework for categorizing flexibility is based on the persistence length $L_p$, measuring the length over which bond orientations remain correlated, and contour length $L_c$ of a polymer.[8, 9] At one end of the spectrum, fully flexible polymers with $L_p \ll L_c$ are conceptualized as freely-jointed chains exhibiting a random coil conformation with negligible bending stiffness. The orientations of monomer subunits are completely uncorrelated. Typical examples include high-molecular-weight, synthetic polymers such as polyethylene, polyamides, and polyesters, whose chains readily adopt random-coil conformations in solution.[6] At the opposite end of the spectrum, rod-like polymers, such as poly(p-phenylene terephthalamide) (Kevlar) and poly(p-phenylene benzobisoxazole) (PBO), consist of rigid, unbending chains.[7] Semiflexible polymers lie in the middle of the spectrum, with their persistence length comparable to the contour length. This intermediate stiffness gives rise to unique behaviors that distinguish semiflexible chains from either flexible or rigid chains.[10-12]

Semiflexible polymers are particularly important for biological systems. Many double-stranded DNA structures, (poly)peptides, and polysaccharides possess characteristic persistence lengths that place them in the semiflexible regime, and their rigidity is crucial for their biological roles.[13, 14] For example, the semiflexible nature of actin filaments allows them to provide mechanical support and enable controlled deformation within the cytoskeleton, which is essential for cell motility and shape changes.[15] Beyond natural systems, semiflexible polymers have also been utilized in various biomedical applications, including drug delivery and biomedical applications.[16, 17] Polymers such as chitosan and hyaluronic acid are widely used as carriers in drug delivery due to their suitable stiffness and biocompatibility,[17] while polymers like polyaniline and water-soluble poly(isocyanopeptides) are utilized in biosensing and cell culture applications.[16]

An important feature of semiflexible polymers is their sensitivity to confinement.[18] In biological systems, DNA encapsulated within viral capsids adopts tightly packed, highly ordered states due to the combined effects of chain stiffness and strong confinement.[14, 19] Similarly, actin filaments confined within vesicles or cells exhibit ordering that contributes to cellular stability and dynamic remodeling.[20-24] In technological contexts, semiflexible polymers in evaporating aerosol droplets can undergo self-assembly under the evolving confined environment.[25] Confinement restricts the accessible conformations of polymers, induces orientational order, and can lead to structural transitions. For semiflexible polymers, confinement affects both enthalpy and entropy of the system. The competition between these contributions governs whether the chains remain isotopically distributed, align into nematic arrangements, or form more complex ordered morphologies.[26]

Given its fundamental and technological importance, the research on the assembly of confined semiflexible polymers has gained traction, with many studies aiming at understanding how confinement influences their assembly, orientational order, and structural transitions.[9, 27-29] Specific focus has been given to the behaviors subject to spherical confinements, including on two-dimensional (2D) surfaces[27, 30] and within three-dimensional (3D) cavities and shells,[19, 28, 31-35] which serve as conceptual models of real confinements. Numerous theoretical and simulation studies have demonstrated diverse structural arrangements, ranging from isotropic and nematic-like states to ordered surface patterns, including bipolar, quadrupolar, tennis-ball, and helical configurations.[28, 30, 36] The emergence of these states is governed by a complex interplay between polymer properties, including $L_c$ and $L_p$, and confinement parameters such as cavity size and surface curvature. Most studies have focused on the behavior of individual semiflexible polymer chains under confinement, providing fundamental insights into their structures,[19, 27, 30-32] while fewer have explored densely packed chains, where the collective effects give rise to even richer structural organization and complex defect patterns.[33, 36]

The confinement can be dynamic, constantly changing shape and even size, as exemplified by evaporating aerosol droplets or the cytoplasm of a living cell. Despite this, in the majority of previous studies, the confinement is treated as rigid and fixed in shape,[30, 33, 35, 36] and how polymers assemble within a deformable and dynamic confinement remains elusive. Here, we employ dissipative particle dynamics (DPD) simulations to investigate the self-assembly of semiflexible polymers confined within deformable droplets, which may also undergo size changes (equivalent to the degree of confinement). Our focus is to elucidate how polymer length and concentration govern structural organization. We uncover distinct regimes ranging from disordered states to various ordered, fibrillar structures. These findings contribute to a deeper fundamental understanding of semiflexible polymer assembly and mechanics in dynamic confinements.

## Methods

### Dissipative Particle Dynamics

Capturing the assembly structure of semiflexible polymers in a droplet, which is a dense and deformable system, requires simulation methods that balance accuracy with computational efficiency. While coarse-grained molecular dynamics and Monte Carlo approaches have been employed extensively to investigate similar systems,[10, 37, 38] DPD offers a particularly powerful alternative.[39, 40] As a coarse-grained method, DPD enables mesoscale simulations of polymer dynamics while retaining essential hydrodynamic interactions and thermal fluctuations. These features make DPD especially well-suited for studying semiflexible polymer assembly in droplets, as it can simultaneously capture emerging behaviors of polymer chains while accurately simulating the hydrodynamics of droplet deformation.

DPD uses coarse-grained beads to represent clusters of atoms, molecules, or polymer segments. Each bead therefore embodies both the mass and interactions of multiple underlying molecular units. This coarse-grained representation dramatically reduces the number of particles required to describe the system, enabling access to much larger system sizes and longer simulation timescales than would otherwise be computationally feasible. The motion of each bead $i$ is determined by Newtonian mechanics:

$$\frac{d\boldsymbol{r}_i}{dt} = \boldsymbol{v}_i \text{ and } m_i \frac{d\boldsymbol{v}_i}{dt} = \boldsymbol{f}_i \tag{1}$$

The total force acting on a bead is composed of several contributions, each representing a different physical mechanism

$$\boldsymbol{f}_i = \sum_{j \neq i} \boldsymbol{F}_{ij}^C + \boldsymbol{F}_{ij}^D + \boldsymbol{F}_{ij}^R + \sum_{j \neq i} \boldsymbol{F}_{ij}^B \tag{2}$$

The conservative interaction, $\boldsymbol{F}_{ij}^C$, provides a soft repulsion between beads

$$\boldsymbol{F}_{ij}^C = a_{ij}\left(1 - \frac{r_{ij}}{r_c}\right)\hat{\boldsymbol{r}}_{ij} \tag{3}$$

where $a_{ij}$ is the DPD interaction parameter. The magnitude of this repulsion depends on the bead types. This parameterization makes it possible to capture macroscopic material properties in a mesoscopic simulation. $r_c$ is the cutoff radius of this interaction, $r_{ij} = \left|\boldsymbol{r}_i - \boldsymbol{r}_j\right|$ and $\hat{\boldsymbol{r}}_{ij} = \left(\boldsymbol{r}_i - \boldsymbol{r}_j\right)/r_{ij}$. In addition to conservative interactions, DPD includes a dissipative force that damps relative motion between neighboring beads. This term serves as a viscous drag, representing the resistance to flow in the coarse-grained fluid

$$\boldsymbol{F}_{ij}^D = -\gamma\left(1 - \frac{r_{ij}}{r_c}\right)^2 \left(\boldsymbol{r}_{ij} \cdot \boldsymbol{v}_{ij}\right)\hat{\boldsymbol{r}}_{ij} \tag{4}$$

where $v_{ij} = \left|\boldsymbol{v}_i - \boldsymbol{v}_j\right|$ is the relative velocity between the beads, and $\gamma$ is the friction coefficient. Complementing the dissipative force is a random force that introduces thermal fluctuations into the system

$$\boldsymbol{F}_{ij}^R = \sigma\left(1 - \frac{r_{ij}}{r_c}\right)\Theta(\Delta t)^{-\frac{1}{2}}\hat{\boldsymbol{r}}_{ij} \tag{5}$$

where $\sigma$ is noise strength, and $\Theta$ is a Gaussian random number with zero mean and unit variance. The dissipative and random forces are coupled through the fluctuation-dissipation theorem, which ensures that the system maintains the correct equilibrium temperature. In the expression for the random force, the factor $(\Delta t)^{-\frac{1}{2}}$ ensures that the stochastic contribution has the correct variance under the fluctuation-dissipation theorem. This scaling maintains the proper balance between the random and dissipative forces as the integration time step $\Delta t$ changes. Together, these two forces function analogously to a thermostat, preserving constant temperature conditions while allowing for realistic hydrodynamic behavior.

Finally, for polymer systems, bonded interactions are included to enforce connectivity between successive oligomer subunits along the chain. These bonded forces are typically harmonic springs that keep beads at an equilibrium separation distance while still allowing conformational flexibility

$$\boldsymbol{F}_{ij}^{B} = -k\left(\boldsymbol{r}_{ij} - b\right)\hat{\boldsymbol{r}}_{ij} \tag{6}$$

By choosing appropriate spring constants, $k$, and equilibrium bond length, $b$, we can model chain connectivity. Chain stiffness is controlled through an angular potential

$$E_a = K[1 + \cos(\beta)] \tag{7}$$

where $K$ is the parameter that tunes the stiffness of the chains and $\beta$ is the angle between consecutive bonds. In this model, the polymer contour length is $L_c = (N-1)b$ with $N$ being the number of polymer beads in a chain and the persistence length is $L_p \approx bK$ when $K \geq 2$.[41, 42]

**Simulation System**

Polymers are modeled as linear chains of coarse-grained beads connected by harmonic springs. The bond stiffness, angle coefficient, and equilibrium bond length are chosen to control the flexibility of the backbone while ensuring numerical stability. These parameters were carefully selected based on previous studies to prevent unphysical bond-crossing, ensuring that the dynamics of polymer chains is physical.[43-45] At the beginning of the simulations, polymer chains are placed within the droplet in random-coil conformations and then solvated with a theta solvent. This initialization ensures that the system does not contain any artificial ordering. The starting configuration consists of a spherical droplet with a radius of 20 in reduced units. The droplet composition is set to 50% polymer beads and 50% solvent beads for the reference system. The droplet is placed in a simulation box filled with a second fluid that is immiscible with solvent and polymer, which maintains the external pressure and prevents artificial compression or expansion of the droplet. The average particle number density of the system is 3 in DPD units. Periodic boundary conditions are applied in all three directions to eliminate edge effects and to mimic a bulk environment. Following the standard practice in DPD, simulations are performed in reduced units. All simulations are conducted using LAMMPS, an open-source molecular dynamics package. The detailed parameter values employed in this study are provided in Table 1.

Table 1. Key parameters used in the dissipative particle dynamics (DPD) simulations.

| **Category** | **Parameter** | **Value** |
|---|---|---|
| **Basic simulation settings** | Units | DPD reduced units |
| | Boundary conditions | Periodic in all directions |
| **DPD conservative parameters ($a_{ij}$)** | Second fluid-Second fluid | 25 |
| | Second fluid-Solvent | 100 |
| | Second fluid-Polymer | 100 |
| | Solvent-Solvent | 25 |
| | Solvent-Polymer | 25 |
| **Bonded interactions** | Bond style | Harmonic |
| | Bond coefficient | $k = 64, r_0 = 0.5$ |
| **Angular interactions** | Angle style | Cosine |
| | Angle coefficient | $K = 50$ |
| **Time** | Timestep | 0.01 |
| **Run duration** | Production runs | 7,000,000 steps |

**Modeling Dynamic Confinement**

An essential aspect of this study is the dynamic evolution of confinement, which is implemented by homogeneously removing solvent beads from the droplet to change its size (degree of confinement). This leads to a linear decrease in confinement volume, which may represent cell volume regulation in specific cellular processes.[46] Nevertheless, this confinement evolution mode should be interpreted as an idealized protocol for simulation simplicity, rather than a direct representation of osmotic or evaporative processes in physical/living systems. Throughout our study, the removal rate, defined as the number of beads removed per physical simulation time, is denoted as $N_{sr}$. When all solvent beads are removed, the droplet is considered to have reached its final, solidified state. To avoid density fluctuations and ensure incompressibility of the surrounding medium,[47] the second fluid beads are introduced simultaneously near the boundaries of the simulation box during the removal process.[48] Notably, several alternative protocols have been employed in DPD to model solvent removal and confinement changes. These include modulating the interactions between solvent and surrounding fluid to drive evaporation at controlled rates,[49] periodically converting selected solvent beads into non-solvent beads within an evaporation zone,[50] and a moving interface method where the droplet boundary recedes and particles near the interface experience a confining force.[51]

**Nematic Order Calculation and Identification of Polymer Bundles and Chain Relative Extension**

To quantify the orientational ordering of the polymer chains, we compute the pairwise nematic order parameter [52, 53]

$$S = \langle \frac{3\cos^2\theta_{ij} - 1}{2} \rangle_{i \neq j} \tag{8}$$

with $\theta_{ij}$ being the angle between two bond vectors $i$ and $j$. The order parameter is obtained as the ensemble average over all pairs of bond vectors separated by less than a cutoff distance. The global nematic order parameter for the entire system is calculated as the mean value over all such bond pairs. Figure S1 shows how the choice of cutoff distance affects the computed nematic order parameter. We note that while the absolute values of the nematic order parameter depend on the choice of cutoff distance, the relative trend remains unaffected.

Polymer bundles are identified by grouping polymer chains according to the spatial proximity of their centers. The center position of each chain is computed by averaging the coordinates of all beads belonging to that chain, which is then used as the input for a density-based clustering procedure to identify groups of chains that are physically assembled into bundles. The density-based spatial clustering algorithm (DBSCAN) is employed here because it does not require specifying the number of bundles in advance and is well-suited for detecting clusters of arbitrary shape.[54] The algorithm assigns chains to the same bundle when they lie within a specified neighborhood radius and when a minimum number of chains are mutually reachable through neighboring chain centers. Chains that do not satisfy these criteria are classified as noise and treated as isolated chains rather than members of any bundle. Each cluster produced by the algorithm is interpreted as a distinct polymer bundle. For visualization and further analysis, all chains within a bundle are grouped together and displayed in a single color.

In addition to identifying polymer bundles, the relative extension of each polymer chain is also quantified.[55, 56] For every chain, the end-to-end distance is calculated from the displacement between the first and last beads in the chain. A dimensionless relative extension ratio is then defined as the end-to-end distance divided by the contour length. Chains with a relative extension ratio above a prescribed threshold of 0.8 are classified as extended chains. The percentage of extended chains in the system is obtained by comparing the number of chains exceeding this threshold to the total chain count. We note that varying the threshold value slightly does not alter the qualitative trends observed in the fraction of extended chains, similar to the insensitivity of the results to the cutoff distance used in bundle identification.

**Results and Discussion**

We perform a series of DPD simulations to quantitatively characterize the assembly behavior of semiflexible polymers confined within a droplet. In the first part of the study, we focus on how the polymer concentration influences the structural organization of the confined system while keeping both the polymer chain contour length and rigidity fixed at $L_c = 25$ and $K = 50$ respectively. The droplet radius is 20 in all of our simulations. As shown in Fig. 1, by systematically varying the polymer concentration, we observe a monotonic increase in the global nematic order parameter, $\langle S \rangle$, indicating that the degree of ordering among the polymer chains becomes progressively stronger as the system becomes more crowded. $\langle S \rangle$ increases by approximately a factor of five when the polymer concentration rises from 10% to 50%. This trend reflects that, at higher concentrations, the polymers align more uniformly, forming locally ordered fibrillar domains, whereas at lower concentrations, the chains remain largely disordered with weak orientational correlations. The emergence of fibril-like structures can be attributed to competition between interchain interactions and the crowding effect under confinement. As the available space decreases, the polymers minimize the system's free energy, in particular bending energy, by aligning their backbones, thereby promoting the formation of parallel fibrillar aggregates.[25, 57, 58] Such behavior highlights the important role of concentration in driving the transition from a disordered to an ordered phase in confined semiflexible polymer systems.

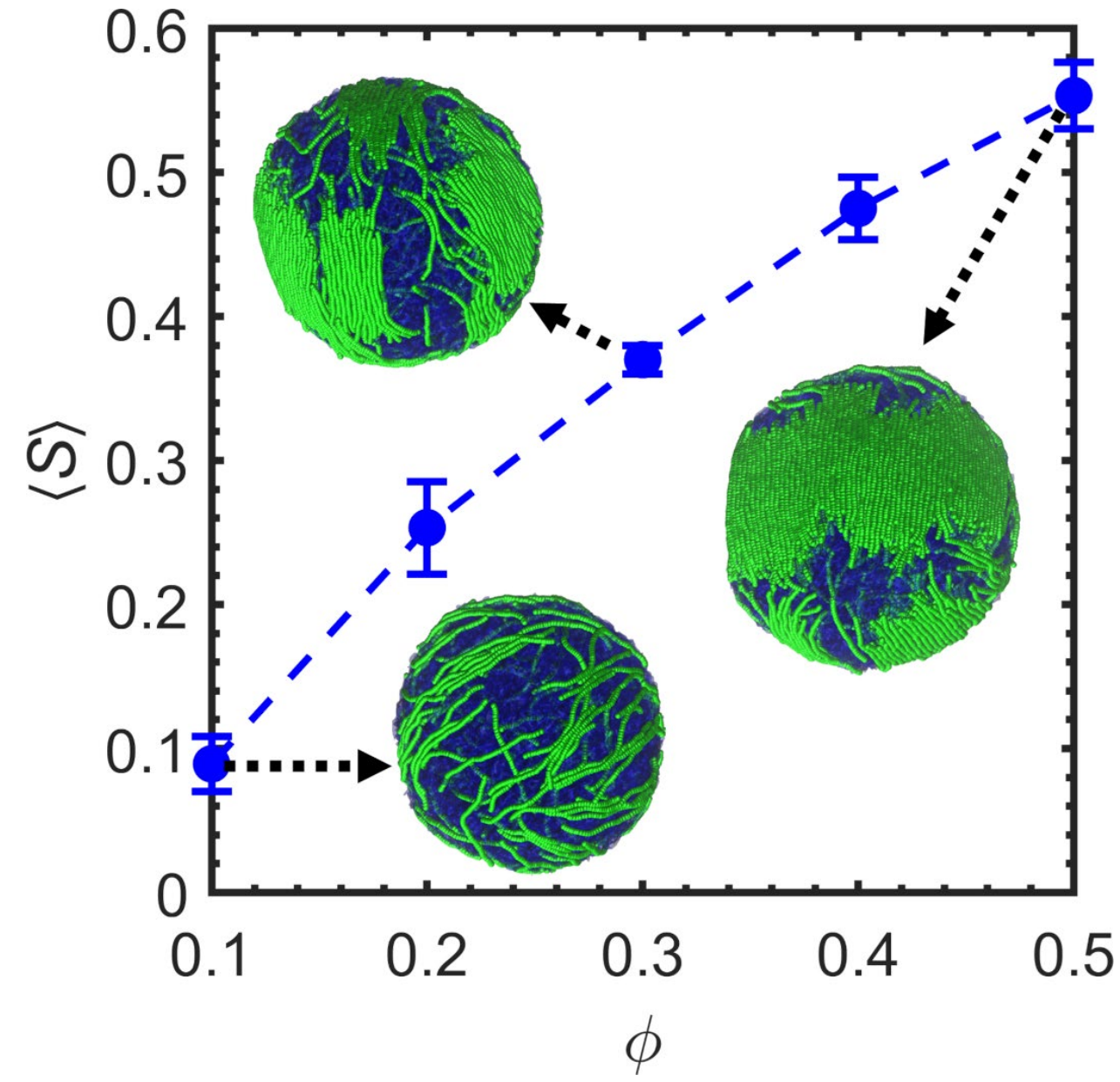

Figure 1. Global nematic order parameter of polymer chains as a function of increasing starting concentration for a fixed contour length of 25. Error bars represent the standard deviation obtained from three independent runs. Insets are representative simulation snapshots showing the final polymer assemblies for different concentrations. Green spheres represent polymer beads. The volume occupied by the solvent is displayed as transparent blue surfaces.

Furthermore, our simulations show that structural ordering develops progressively over time, initiating at the droplet surface and subsequently propagating toward the droplet interior (see Video S1). The video also clearly indicates that the chains preferentially form ordered regions in the proximity of the droplet surface. To illustrate this behavior, spatial variations in chain ordering are characterized in the $\phi = 0.1$ and $0.5$ systems and presented in Fig. S2. In both systems, the local nematic order parameter is significantly higher near the droplet interface than at the center. For example, in the 50% polymer system, regions close to the interface exhibit approximately six times higher local ordering than those near the center of the droplet. This heterogeneous ordering is further supported by the cross-sectional simulation snapshots of the droplet (Fig. S2 insets). The chains near the interface are markedly more ordered than those in the interior regions. This behavior arises because the interface acts as a geometric template that promotes local chain alignment. The surface imposes spatial constraints and orientational bias, favoring alignment of polymer backbones parallel to the interface, which facilitates the nucleation of ordered fibrillar domains.[37] This enhanced ordering in the interfacial regions suggests that the droplet surface plays a critical role in promoting chain alignment and structural organization within the system.

To investigate the dynamic evolution of assembly and resulting heterogeneity, we analyze the radial distribution of polymer density. As shown in Fig. 2a, the polymer density is relatively uniform throughout the droplet at the beginning of the simulation. As the simulation progresses, polymer density near the droplet surface gradually increases, indicating an accumulation of chains in the interfacial region. To further confirm this trend, Fig. S3 shows that the number of polymer chains in the interior regions decreases over time, while simultaneously increasing in regions closer to the droplet surface. This result clearly demonstrates the strong correlation between chain density and chain ordering. Figure 2b presents the equilibrium radial polymer density distributions for systems with different polymer concentrations. As it shows, at lower concentrations, the polymer density remains nearly uniform across the droplet, implying a disordered configuration. In contrast, as the polymer concentration increases, a pronounced peak develops near the surface region, revealing that higher concentrations enhance polymer accumulation. Notably, the maximum density in the profile appears not exactly at the interface but slightly interior to it. This behavior could be attributed to anisotropic droplet deformation induced by fibrillar assembly and to surface curvature that frustrates further domain growth.

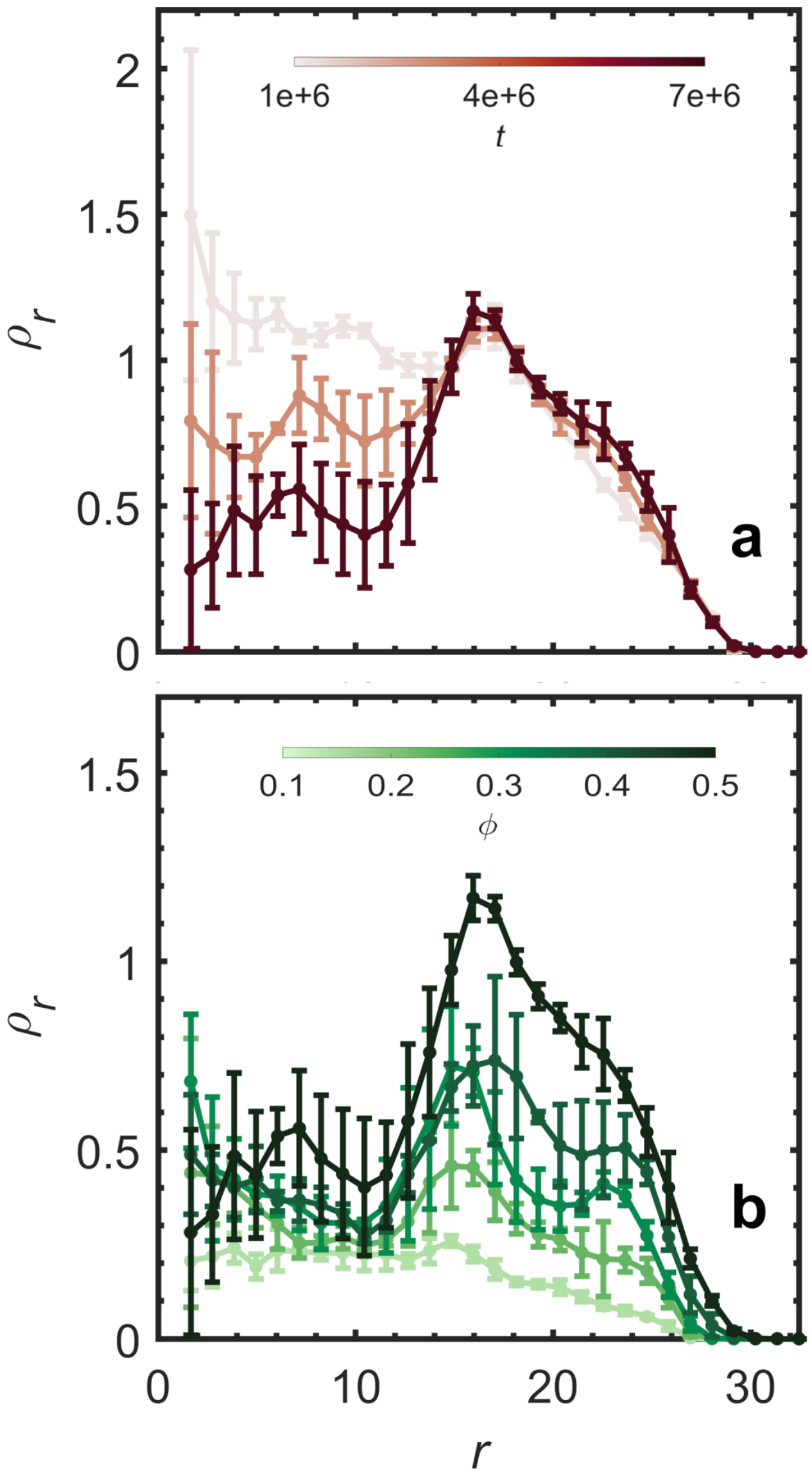

Figure 2. Radial distribution of polymer density for systems with a polymer length of 25. (a) Time evolution of the average radial polymer density distribution for a system with a polymer concentration of 0.5. Different shades of red indicate successive time steps, highlighting the temporal evolution of polymer distribution. (b) Equilibrium average radial polymer density distribution for different polymer concentrations. Different shades of green represent varying concentrations, illustrating the effects of concentration on polymer spatial organization. In both subpanels, error bars represent the standard deviations obtained from three independent runs.

To further examine the effect of concentration on the degree of ordering within the system, we also calculate the eigenvalues of the nematic tensor of bond orientation ($Q_{\alpha\beta} = \frac{1}{2}\langle 3u_{ni}^{\alpha}u_{ni}^{\beta} - \delta_{\alpha\beta}\rangle$).[36] In a completely disordered state, all three eigenvalues of the $\mathbf{Q}$ tensor are equal to zero, whereas deviation from zero indicates the development of orientational order. Figure 3 demonstrates that the eigenvalues remain close to zero at lower polymer concentrations, confirming that the system is largely disordered. With increasing concentrations, the eigenvalues become more distinct and deviate from zero, reflecting the emergence of orientational alignment among the polymer chains and the overall transition toward an ordered structure. This trend is consistent with the findings previously reported by Nikoubashman et al.[36]

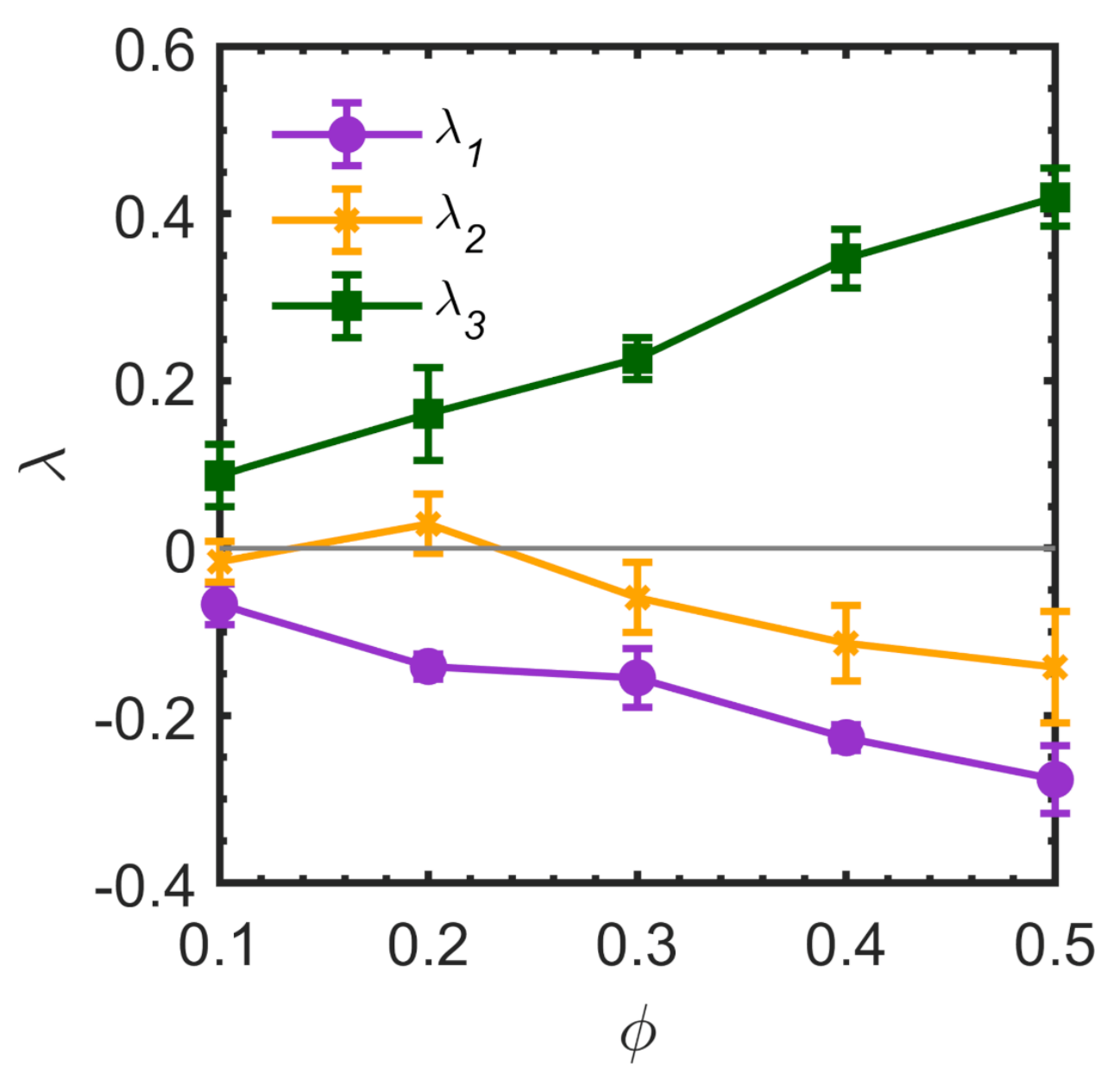

Figure 3. Eigenvalues of the nematic tensor $\mathbf{Q}$ for systems with a polymer length of 25 at different concentrations. The plot shows the three eigenvalues where $\lambda_1 \leq \lambda_2 \leq \lambda_3$ as a function of concentration, providing insight into the degree and nature of orientational ordering within the polymer assemblies. Error bars represent the standard deviations obtained from three independent runs.

As the next part of our study, we explore the assembly structure of polymers as a function of increasing chain contour length while keeping the polymer concentration fixed at 0.5. Representative snapshots of the system in Fig. 4 reveal three distinct types of structural organization. For shorter polymer chains, the system remains largely disordered, with polymer chains exhibiting random orientations and minimal local alignment. At intermediate chain lengths, we observe the formation of linear polymer fibrils, indicating the emergence of local orientational order and alignment along preferential directions. For the longest chains, circular polymer bundles form, reflecting the development of a more complex, curved assembly structure. The distinction among these three regimes is also reflected in the trend of the global nematic order parameter. As the polymer length increases, the nematic order parameter initially rises, reaching a maximum when the system is dominated by linear fibrillar domains, and then decreases as circular bundle domains become prevalent. These results demonstrate that the polymer chain length strongly influences the type and degree of ordering within the droplet. The radial polymer density profiles for the intermediate and longest chain lengths in Fig. S4 confirm again enhanced polymer accumulation near the droplet surface region. Interestingly, for the shortest chain length of 10, the polymer density remains more uniform throughout the droplet. This behavior for the shortest chain length may arise because interfacial adsorption of such short chains leads to a significant reduction in the system entropy, while the corresponding enthalpic gain is insufficient to compensate for this loss. Namely, adsorption at the interface requires linear polymer chains to adopt the local geodesic curvature of the droplet surface, resulting in a considerable bending energy penalty for short and rigid chains. The uniform distribution of short chains even at high concentration also rules out surface tension reduction as a mechanism for interfacial adsorption and the ensuing assembly.

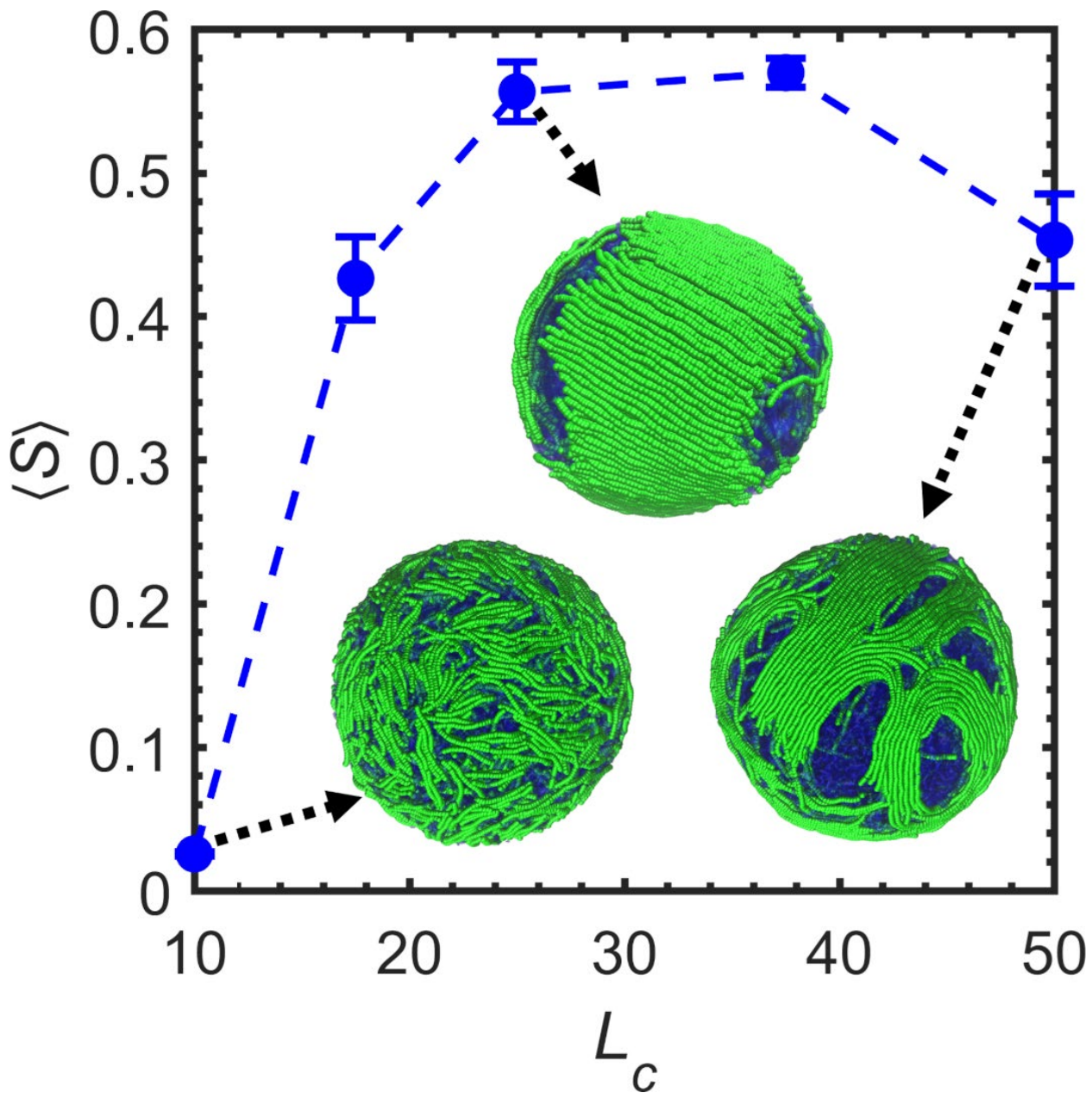

Figure 4. Global nematic order parameter of polymers as a function of increasing chain length for a fixed concentration of 0.5. Error bars indicate the standard deviations among three independent runs. Insets show representative final polymer assemblies for different chain lengths.

To better visualize the structure of the different polymer ordered domains, we apply the DBSCAN clustering algorithm to identify chain groups based on their spatial proximity and alignment. DBSCAN detects clusters by locating regions of high density separated by low-density boundaries, allowing us to distinguish well-defined ordered domains from isolated or unassociated chains. Detailed algorithmic settings and parameter selections are provided in the Methods section. As shown in Figure 5, the results of this analysis indicate that for the short-chain system, very few ordered bundles are observed, reflecting the largely disordered configuration. In the intermediate-chain system, multiple linear fibrils are clearly visible, indicating enhanced alignment and ordering. For the long-chain system, several circular bundles form, demonstrating the emergence of complex curved structures.

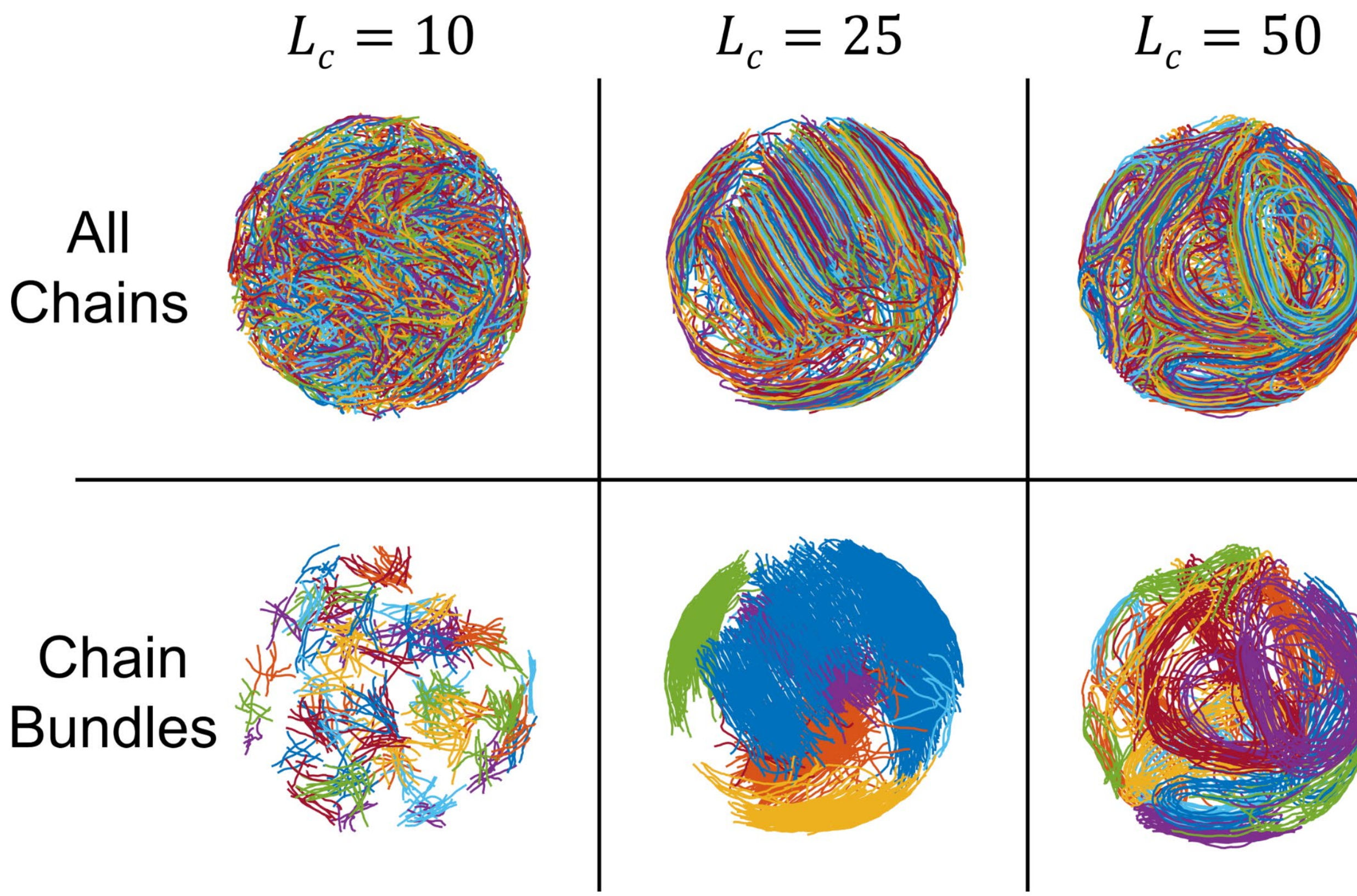

Figure 5. Effect of polymer chain length on bundle morphology. The top row shows snapshots of all polymer chains in the system at various chain lengths. The bottom row highlights individual polymer bundles identified using a proximity-based clustering algorithm, colored distinctly to reveal the structural transition from linear bundles to circular bundle formations as chain length increases.

Through examining the effect of chain length on ordered domains (Fig. 6), we find that the trend in the fraction of polymer chains belonging to ordered domains closely mirrors that observed in Fig. 4 for the nematic order parameter. Initially, as the chain length increases, the fraction of chains in ordered domains rises, particularly in the regime where linear fibrillar domains form, reflecting enhanced ordering. As the system transitions to the regime of circular domains, a slight decrease in the fraction of chains in ordered domains is observed. The decreasing percentage of extended chains shown in Fig. 6 is likely influenced by two factors. First, the decreasing droplet size relative to chain length geometrically limits chain extension. Second, the persistence length is held constant across these simulations, so increasing the contour length reduces the $L_p/L_c$ ratio. This lowers the effective rigidity of the chains and reduces their tendency to align and form linear bundles.

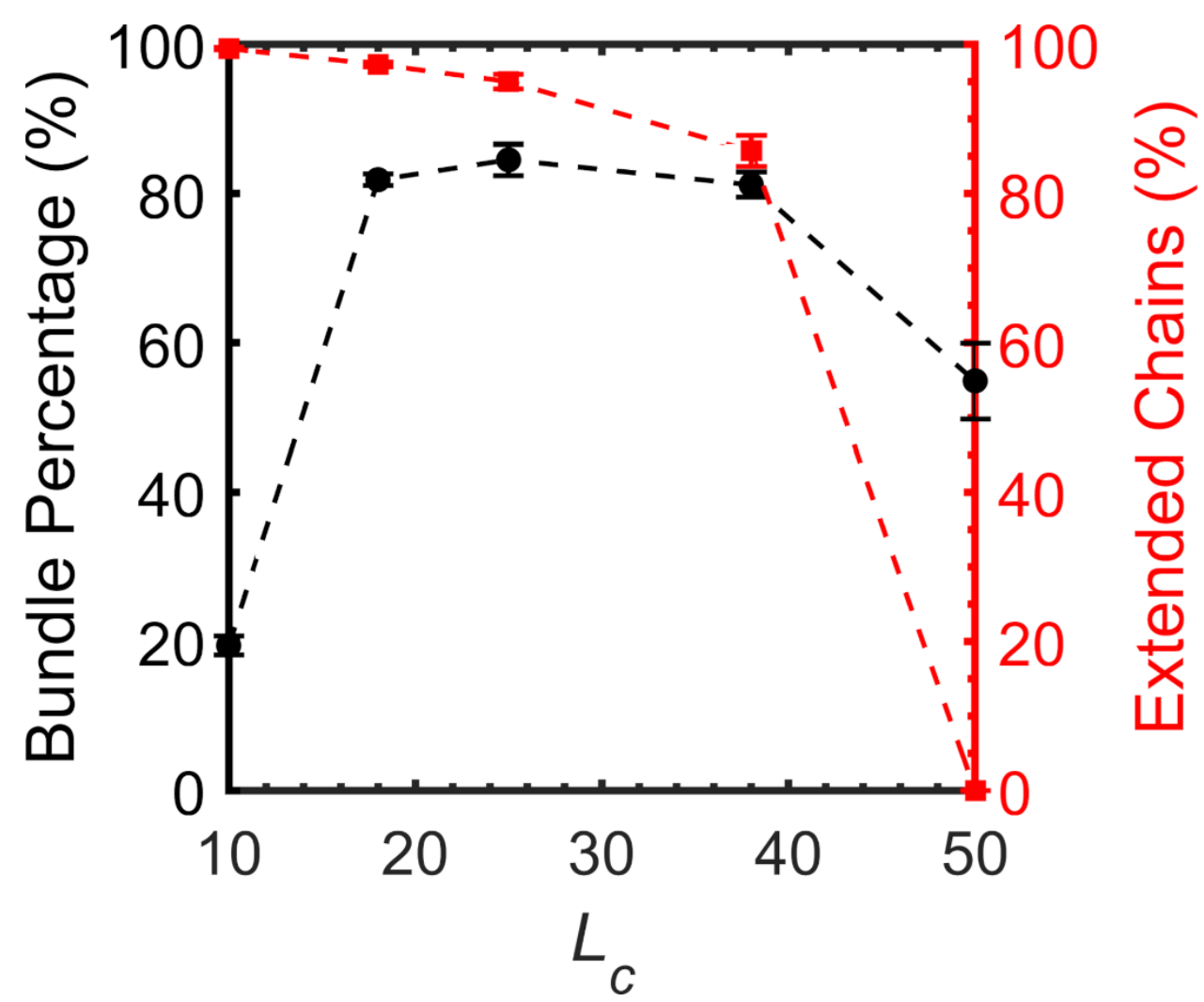

Figure 6. The percentage of polymer chains incorporated into bundles (black), normalized by the total number of chains, and the percentage of polymer chains that fall below the relative extension threshold (red), representing chains with predominantly linear conformations. Error bars represent the standard deviations among three independent runs.

To further investigate the effect of $L_p/L_c$, we also perform an additional simulation for the long-chain system ($L_c = 50$) with a higher bending stiffness ($K = 300$), making the chains even more rigid. As shown in Fig. S5, increasing $L_p/L_c$ leads to considerable changes in the assembly structure: the droplet becomes elongated, and the circular bundles exhibit a larger radius than in the original case with $K = 50$. These results demonstrate that both chain rigidity and confinement influence the final assembly structures.

We further explore the impact of dynamic confinement on polymer assembly. The confinement is systematically controlled by gradually removing solvent beads over time (see Video S2). A key control parameter in this process is the rate at which solvent beads are removed ($N_{sr}$), which determines how rapidly the droplet shrinks and how quickly the polymer concentration increases. As evident from both the quantitative trend and the corresponding snapshots of the final polymer assemblies in Fig. 7, the structure exhibits significantly higher ordering at slower removal rates. In contrast, as the removal rate increases, the degree of ordering progressively decreases, leading to more disordered final configurations. The global nematic order parameter becomes nearly nine times larger at the final stage when the solvent bead removal rate is reduced from the highest to the lowest.

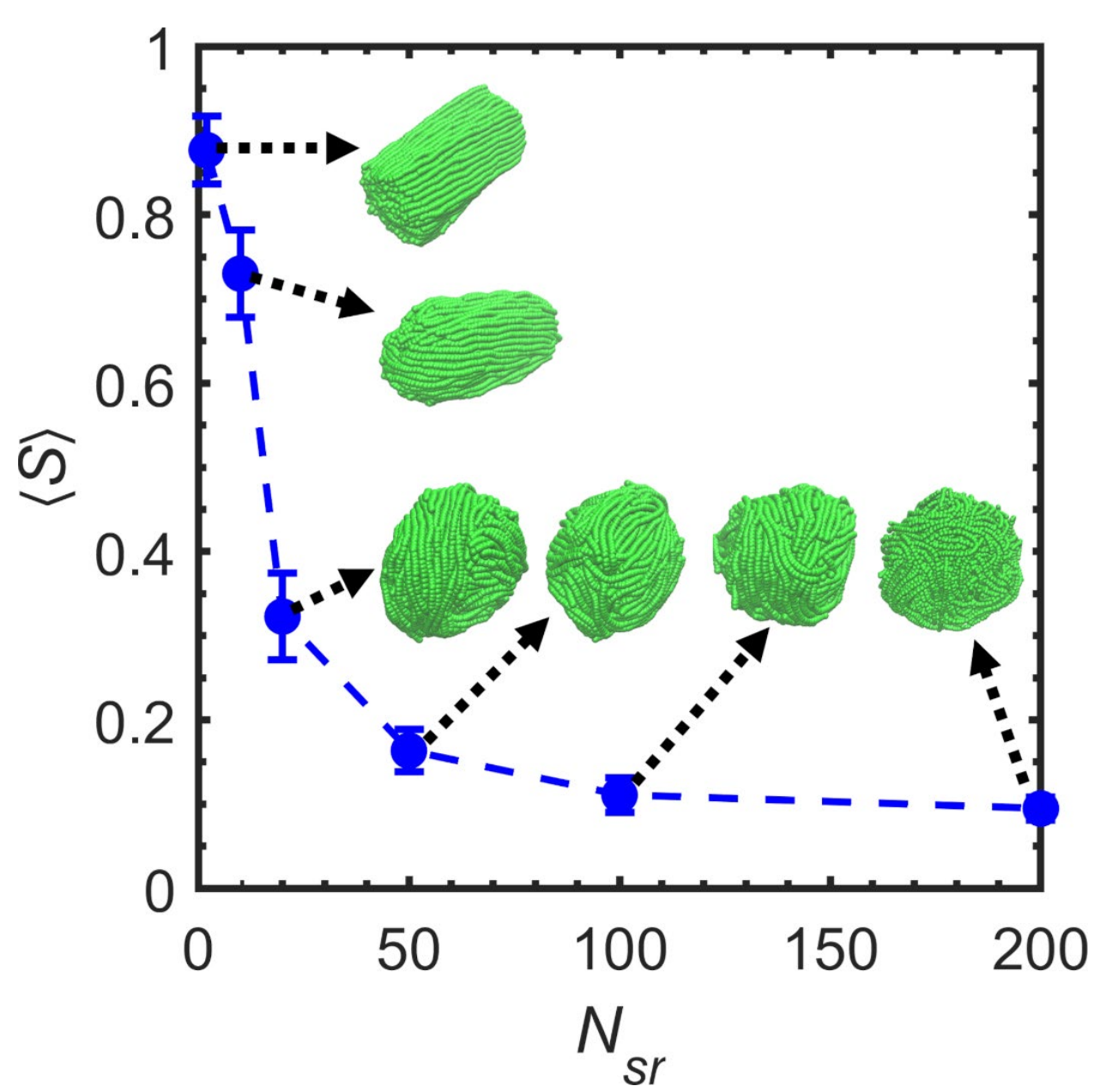

Figure 7. Global nematic order parameter as a function of increasing solvent removal rate for a system with a polymer length of 25 and a concentration of 0.5. Error bars represent the standard deviations among three independent runs. Representative snapshots of the polymer assemblies after solvent removal are shown as insets, illustrating structural differences at different rates.

This behavior arises because, when the solvent beads are removed slowly, the polymer concentration increases gradually, providing sufficient time for the system to reorganize and reach quasi-equilibrium configurations as the confinement strengthens. As demonstrated earlier in Fig. 1, increasing polymer concentration promotes the formation of ordered structures. Therefore, the extended time spent in the high-concentration regime at slow removal rates enables the chains to align and assemble into more ordered morphologies. In contrast, when the solvent beads are removed rapidly, the concentration of polymers also increases, but the process occurs so quickly that the system becomes kinetically arrested before it can equilibrate or nucleate ordered domains. The rapid removal rate effectively quenches the structure, creating a disordered state that frustrates the formation and growth of fibrillar order. This kinetic limitation is consistent with prior observations, where fast solvent evaporation suppresses salt crystallization by trapping the system in a nonequilibrium, amorphous configuration.[59] To further investigate this behavior, we also examine the influence of the initial polymer concentration. Additional simulations are performed by reducing the initial polymer concentration from 0.5 to 0.1 while keeping the solvent removal rate, $N_{sr}$, constant. As shown in Fig. S6, the overall trend remains unchanged, although the system with the lower initial concentration exhibits a slightly higher nematic order parameter. This behavior is attributed to the fact that decreasing polymer concentration increases the amount of solvent in the system, resulting in a longer time for assembly into higher ordered structures at the same $N_{sr}$.

As shown in the snapshots in Figure 7, the differences between the final assembly structures are not limited to the degree of ordering as discussed, but are also reflected in their overall shapes. Specifically, the final assemblies obtained at higher solvent bead removal rates appear more spherical, whereas those formed at lower removal rates deviate more from spherical symmetry and exhibit elongated or anisotropic shapes. This observation suggests that the removal rate not only influences internal structural order but also affects the global morphology of the polymer assembly.

To quantify these differences in shape more precisely, we calculate two shape descriptors based on the gyration tensor, as shown in Figure 8: the relative shape anisotropy ($\kappa^2$) and the asphericity ($b$). $\kappa^2$ is a dimensionless parameter that measures the overall deviation of a structure from an ideal sphere, with $\kappa^2 = 0$ corresponding to a perfectly spherical object and larger values indicating more anisotropic or elongated configurations. $b$ provides complementary information by quantifying the variance among the principal moments of the gyration tensor. Increasing values of $b$ correspond to shapes that progressively lose spherical symmetry. Both metrics consistently show a decreasing trend with increasing solvent bead removal rate, indicating that the final assemblies become more spherical. At slow removal rates, the polymer chains have sufficient time to rearrange and form ordered domains that extend over larger length scales, leading to more anisotropic, non-spherical aggregate shapes. Conversely, at fast removal rates, the rapid confinement limits structural rearrangements, trapping the system in more compact and isotropic configurations that appear nearly spherical. We confirm that these kinetically trapped configurations relax only marginally post complete solvent removal (Video S3).

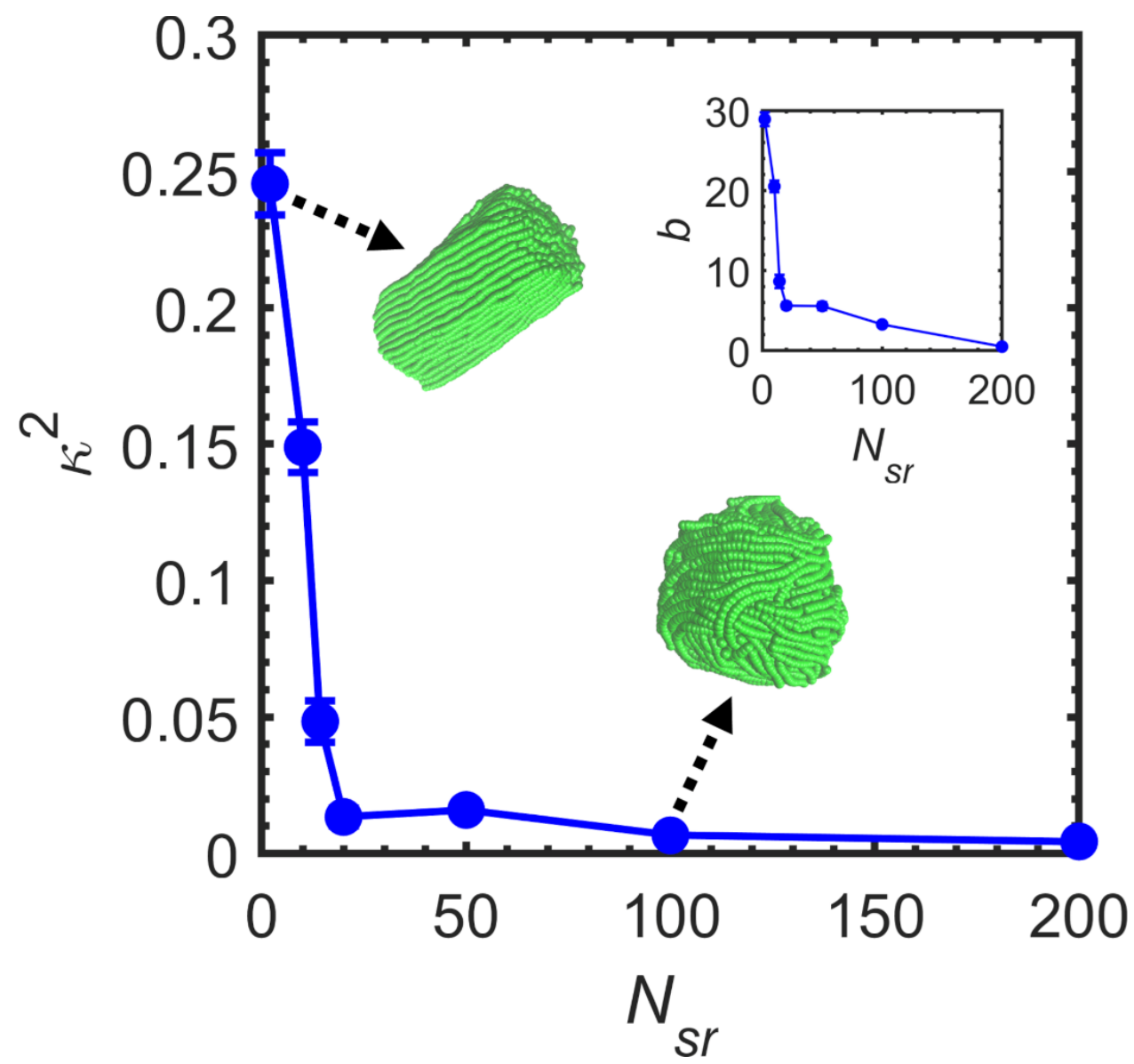

Figure 8. Dependence of relative shape anisotropy of the final assembly on solvent removal rate for a system with a polymer length of 25 and concentration of 0.5. The inset shows the corresponding trend of asphericity. Error bars in both plots indicate the standard deviations among three independent runs.

The combination of visual snapshots and quantitative shape analysis reveals a strong correlation between ordering and shape anisotropy. Slow removal rates promote both higher internal order and more anisotropic, elongated structures, while rapid removal rates yield disordered yet more spherical assemblies. This interplay highlights how kinetic constraints during droplet size change can simultaneously influence the internal structure and overall morphology of polymer assemblies.

## Conclusions

In this work, we employ mesoscopic DPD simulations to systematically investigate the self-assembly of semiflexible polymers confined within a deformable droplet, with particular emphasis on the roles of polymer concentration, chain contour length, and dynamically evolving confinement. By focusing on crowded polymer systems in a nonrigid environment, this study addresses an important gap in the current understanding of polymer assembly under realistic, deformable, and sometimes time-dependent confinement conditions. Our results show that increasing polymer concentration strongly enhances orientational order, driving a transition from disordered conformations to ordered fibrillar structures. This ordering is spatially heterogeneous, with significantly higher ordering near the droplet interface, where the geometric template promotes chain orientation and nucleates ordered domains. Polymer contour length controls the type of assembled structure. Short chains remain largely disordered, intermediate-length chains form linear fibrillar domains with the highest nematic order, and longer chains assemble into circular bundles with reduced global alignment. These trends illustrate how chain stiffness and the increasing flexibility with longer contour lengths influence the resulting polymer assembly structure. Finally, we demonstrate that dynamic confinement critically influences both internal order and global morphology. Slow solvent removal enables structural reorganization and leads to highly ordered, anisotropic assemblies, whereas rapid confinement changes kinetically trap the system in disordered states with nearly spherical shapes. Together, these findings highlight the coupled roles of concentration, chain length, and confinement dynamics in governing the assembly of semiflexible polymers in deformable, dynamic environments, with implications for both biological and technological systems.

**Supporting Information**

Supplementary figures on cutoff distance effect, radial variations in the nematic order, radial density distribution of polymer, bending stiffness effect on long-chain assembly, dynamic confinement system with low polymer concentration; supplementary videos on time evolution of the assembly.

**Acknowledgment**

This research was supported by the U.S. National Science Foundation (NSF) Advanced Manufacturing program through Awards 2335614 and 2335615.

**Supporting Information for**

# Self-Assembly of Crowded Semiflexible Polymers under Dynamic and Deformable Confinement

Nasir Amiri[1], Jonathan P. Singer[2], Xin Yong[1]*

[1]Department of Mechanical and Aerospace Engineering, University at Buffalo, Buffalo, NY 14260, USA

[2]Department of Mechanical and Aerospace Engineering, Rutgers University, Piscataway, NJ 08854, USA

*Email: xinyong@buffalo.edu

## Supplementary Figures

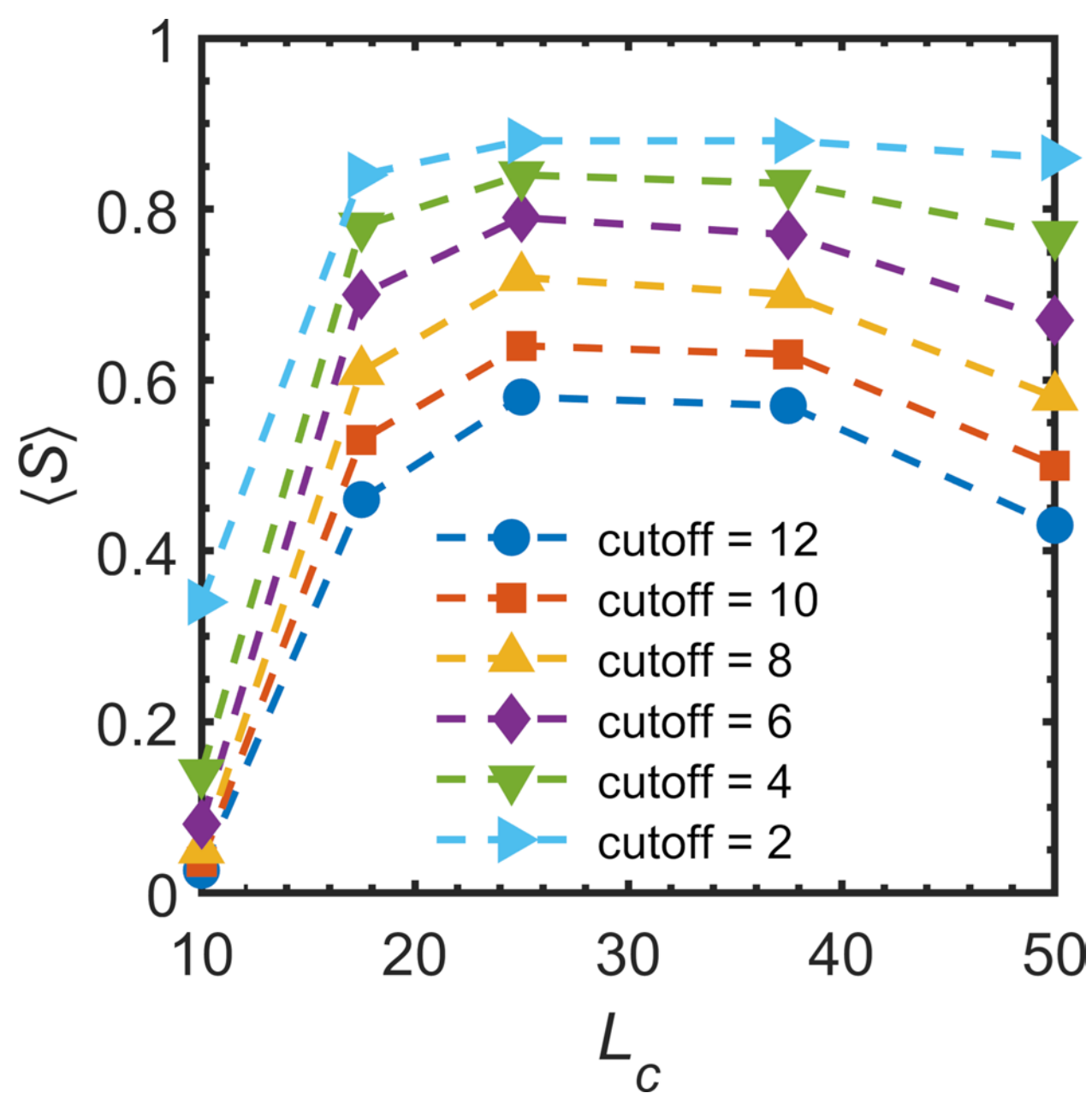

Figure S1. Effects of the cutoff distance on the calculated global nematic order parameter.

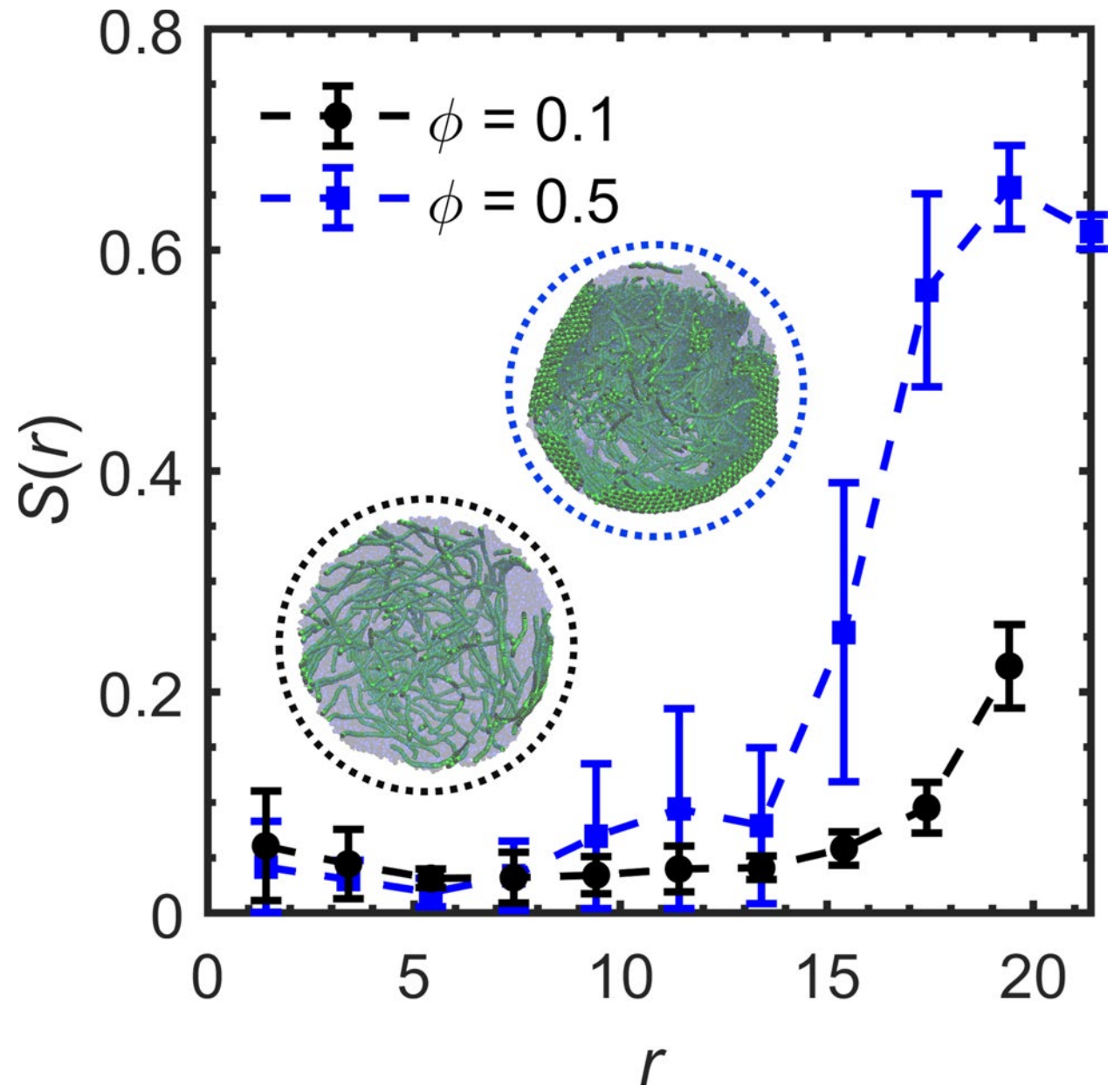

Figure S2. Radial variations in the local polymer nematic order parameter, $S(r)$, where $r$ is measured from the droplet center toward the interface for polymer concentrations of 10% and 50%. Error bars represent the standard deviations among three independent runs. The insets present cross-sectional snapshots of the droplet. The configurations highlighted by the black and blue dotted circles correspond to $\phi = 0.1$ and 0.5, respectively.

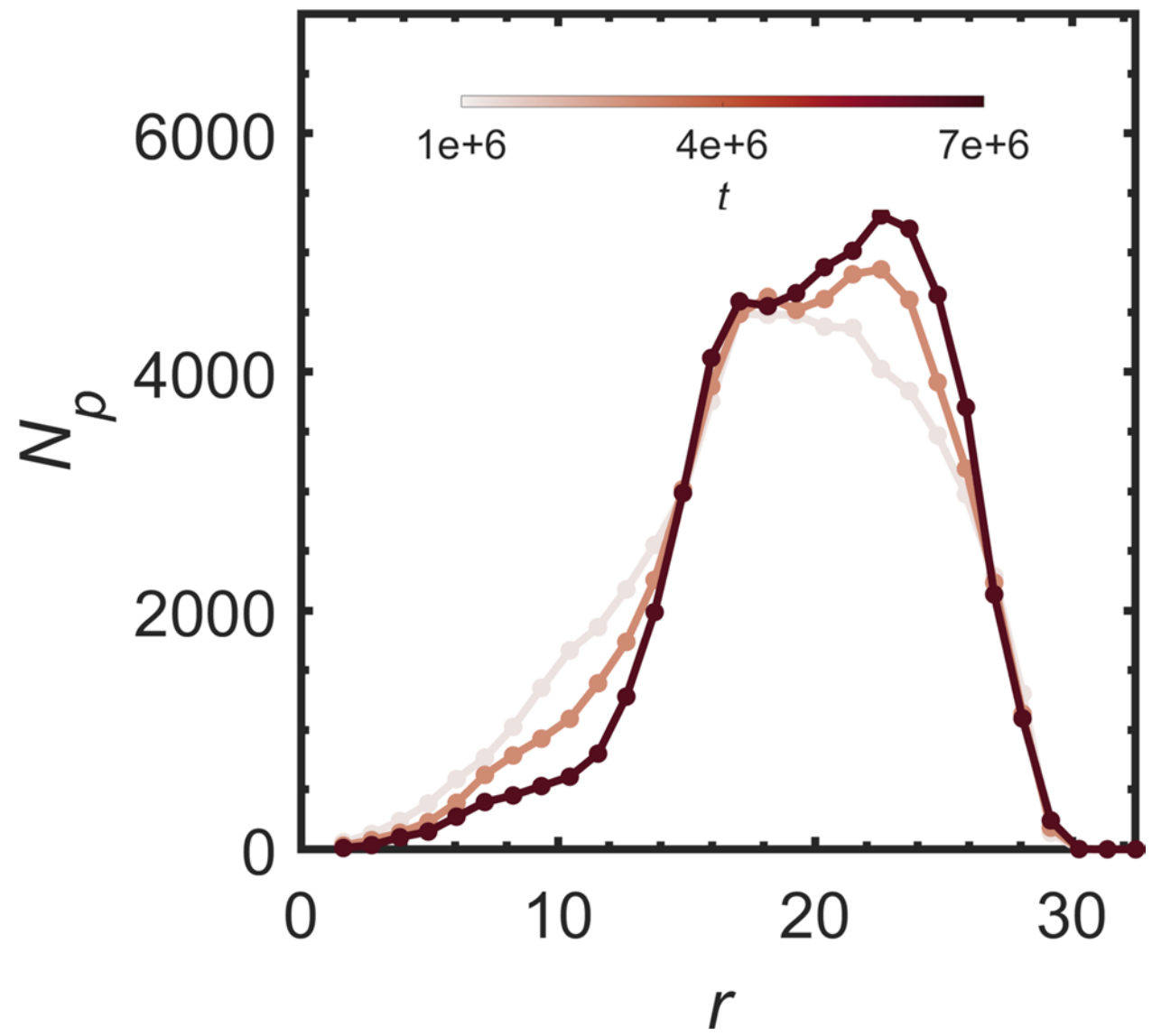

Figure S3. Time evolution of the radial distribution of the number of polymer beads for a system with a polymer concentration of 0.5. Different color intensities represent successive time steps, illustrating the redistribution of polymer chains over time.

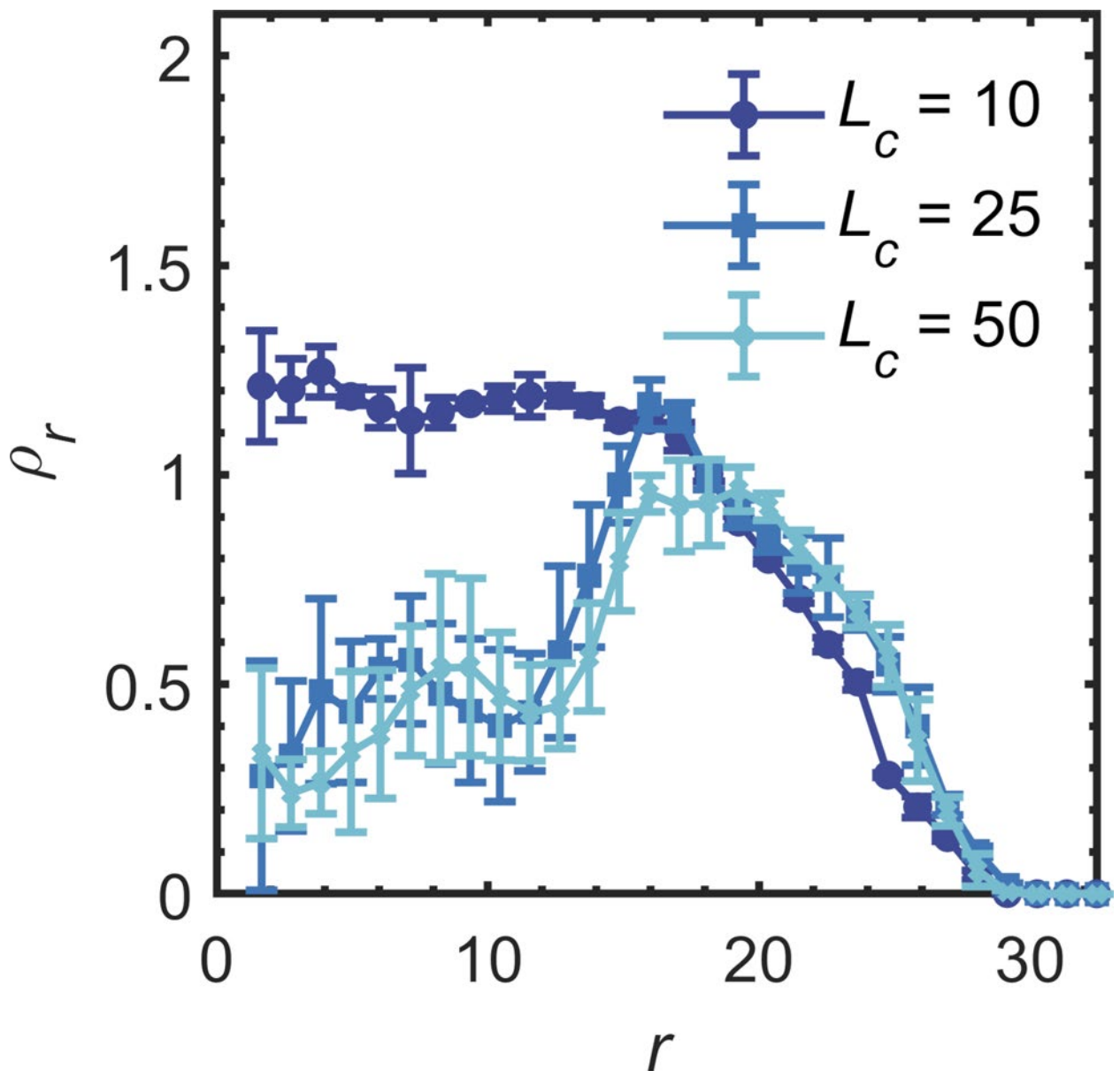

Figure S4. Equilibrium average radial distribution of polymer density for different polymer chain lengths at polymer concentration of 0.5. Error bars represent the standard deviations obtained from three independent runs.

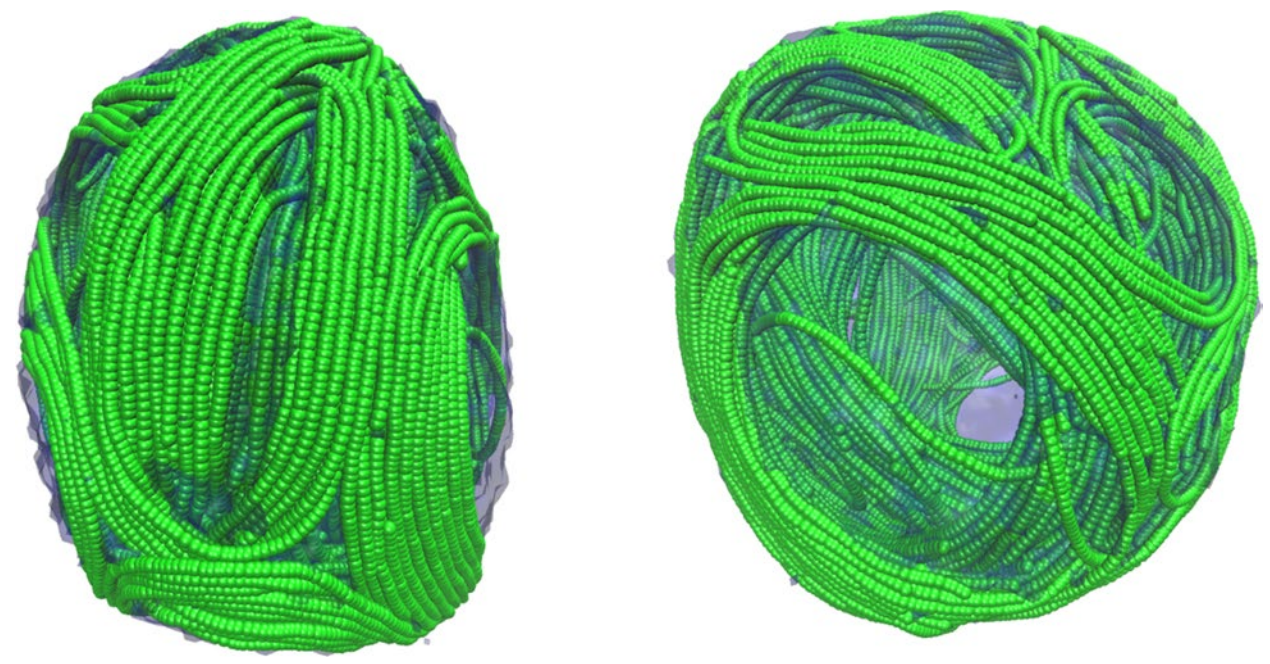

Figure S5. Snapshots of the assembly structure of the long-chain system with high bending stiffness (K = 300) and concentration of 0.5, shown from two different viewing angles.

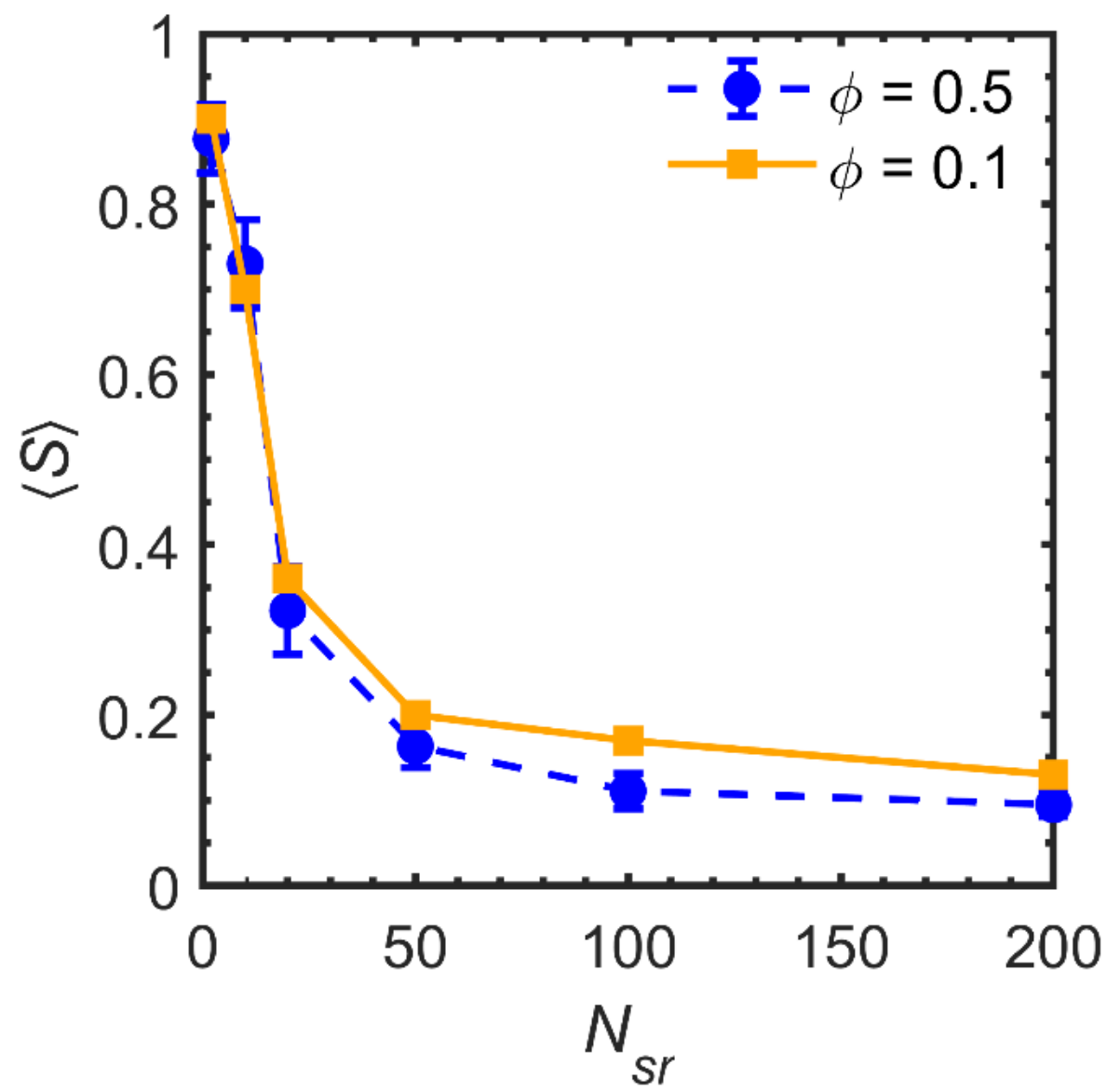

Figure S6. Nematic order parameter as a function of increasing solvent removal rate for systems with a polymer length of 25 and concentration of 0.1 and 0.5.

**Supplementary Videos**

Video S1. Time evolution of the system at a polymer concentration of 0.5 and contour length of 25, shown through cross-sectional snapshots of the droplet. Green spheres represent polymer beads and transparent blue spheres represent solvent beads. The second fluid is not displayed for clarity. The video illustrates the transition from an initially random, disordered configuration to an ordered fibrillar state near the droplet interface as the assembly progresses.

Video S2. Time evolution of a volume-changing droplet at $N_{sr} = 10$, polymer concentration of 0.5, and contour length of 25. The video illustrates the transition from an initially random, disordered configuration to a highly ordered, anisotropic assembly at the end. Green spheres represent polymer beads. The volume occupied by the solvent is displayed as transparent blue surfaces. The second fluid is not displayed for clarity.

Video S3. Extended time evolution of a volume-changing droplet at $N_{sr} = 100$, polymer concentration of 0.5, and contour length of 25. The color scheme is the same as Video S2. The simulation is continued well beyond the completion of solvent removal to monitor the evolution of the assembled polymer structure and assess whether the configuration remains kinetically trapped or relaxes toward equilibrium.